\documentclass[aps,prb,amssymb,twocolumn,showpacs]{revtex4}
\usepackage{graphicx}
\usepackage{times}

\newcommand{\bra}[1]{\langle #1 |}
\newcommand{\ket}[1]{| #1 \rangle}

\newcommand\Nij{{\mathcal N}_{ij}}
\newcommand\Cij{{\mathcal C}_{ij}}
\newcommand\Nt{{\mathcal N}}

\newcommand{\be}{\begin{equation}}
\newcommand{\ee}{\end{equation}}
\newcommand{\bae}{\begin{eqnarray}}
\newcommand{\eae}{\end{eqnarray}}
\newcommand{\ep}{\epsilon}

\def\CC{{\rm\kern.24em \vrule width.04em height1.46ex depth-.07ex
    \kern-.30em C}}
\def\P{{\rm I\kern-.25em P}}

\def\bbbc{{\mathchoice {\setbox0=\hbox{$\displaystyle\rm C$}\hbox{\hbox
to0pt{\kern0.4\wd0\vrule height0.9\ht0\hss}\box0}}
{\setbox0=\hbox{$\textstyle\rm C$}\hbox{\hbox
to0pt{\kern0.4\wd0\vrule height0.9\ht0\hss}\box0}}
{\setbox0=\hbox{$\scriptstyle\rm C$}\hbox{\hbox
to0pt{\kern0.4\wd0\vrule height0.9\ht0\hss}\box0}}
{\setbox0=\hbox{$\scriptscriptstyle\rm C$}\hbox{\hbox
to0pt{\kern0.4\wd0\vrule height0.9\ht0\hss}\box0}}}}

\def\bbbz{{\mathchoice {\hbox{$\sf\textstyle Z\kern-0.4em Z$}}
{\hbox{$\sf\textstyle Z\kern-0.4em Z$}}
{\hbox{$\sf\scriptstyle Z\kern-0.3em Z$}}
{\hbox{$\sf\scriptscriptstyle Z\kern-0.2em Z$}}}}

\begin{document}

\title{Entanglement in extended Hubbard models and quantum phase transitions}

\author{Alberto Anfossi,$^1$ Paolo Giorda,$^{1,2}$ and Arianna Montorsi$^1$}

\affiliation{$^1$Dipartimento di Fisica del Politecnico and CNISM, Corso Duca degli Abruzzi 24, I-10129 Torino, Italy}

\affiliation{$^2$Institute for Scientific Interchange (ISI), Villa Gualino, Viale
Settimio Severo 65, I-10133 Torino, Italy}

\date{15 March 2007}

\begin{abstract}

The role of two-point and multipartite entanglement at quantum phase transitions (QPTs) in correlated electron systems is investigated. We consider a bond-charge extended Hubbard model exactly solvable in one dimension which displays various QPTs, with two (qubit) as well as more (qudit) on-site degrees of freedom involved. The analysis is carried out by means of appropriate measures of bipartite/multipartite quantum correlations. It is found that all transitions ascribed to two-point correlations are characterized by an entanglement range which diverges at the transition points. The exponent coincides with that of the correlation length at the transitions. We introduce the correlation ratio, namely, the ratio of quantum mutual information and single-site entanglement. We show that at $T=0$, it captures the relative role of two-point and multipartite quantum correlations at transition points, generalizing to qudit systems the entanglement ratio. Moreover, a finite value of quantum mutual information between infinitely distant sites is seen to quantify the presence of off-diagonal long-range order induced by multipartite entanglement.
\end{abstract}


\maketitle

\section{introduction}

Among the ground-state properties of many-body quantum systems, correlations are recognized as fundamental to the comprehension of the critical behavior displayed at quantum phase transitions (QPTs).\cite{sadchso} In this respect, in recent years a crucial role has been played by the notion of entanglement between subsystems and the related measures developed within the field of quantum information theory.\cite{qip2}

The large amount of results achieved \cite{fazioNielsen}$-$\cite{CGZ} relies on the observation that a QPT is, in general, characterized through nonanalyticities of the density matrix of an appropriate subsystem. The latter is the starting point for the determination of any measure of entanglement, i.e., quantum correlation, either within the subsystem or between the subsystem and the remaining system. In the latter case, Von Neumann entropy $\mathcal{S}$ is always able to capture the presence of a QPT (of finite order) for an appropriate choice of the subsystem. Also, some general information about the type of transition (for instance, its order) can be gained by looking at the type of singularity in $\mathcal{S}$. \cite{WSL} Nevertheless, in order to construct a complete description of QPTs, one should provide a plethora of other features, such as, for instance, critical exponents. This requires the evaluation of more punctual measures of entanglement, like that of quantum correlations between two points or among more subsystems (multipartite correlations). Such measures have already been investigated in relation to QPTs for qubit systems. In Ref. \onlinecite{fazioNielsen}, it was shown that concurrence (measuring two-point entanglement) scales with universal exponent for an $XY$ model, whereas in Ref. \onlinecite{VerER} the $n$-tangle measure was used to detect the singular behavior of multipartite correlations at a QPT.

A certain number of interesting results have been obtained as well for QPTs and entanglement in {\it correlated electron systems}, \cite{GDLL}$-$\cite{CGZ} the latter being, in principle, characterized by a larger number of degrees of freedom per site (typically 4) with respect to qubit systems. This point makes it necessary to use measures of quantum correlations which are tailored also for qudit systems. Most of such measures are difficult to evaluate whenever the subsystem is in a mixed state, since they require an often out of reach optimization process. In Ref. \onlinecite{AGM}, a method was proposed to distinguish at a given QPT the contribution of two-point entanglement from  that of multipartite quantum correlations without entering the above difficulties. The method is based on the comparative use of single-site Von Neumann entropy $\mathcal{S}_i$ and quantum mutual information $\mathcal{I}_{ij}$; the latter being a measure of all (quantum and classical) correlations between two generic sites $i,j$. The method provides a simple recipe: whenever the two measures display the same type of singularity at a given transition, the latter ascribed to two-point correlations; on the contrary, if the singularity displayed by $\mathcal{S}_i$ is seen differently by $\mathcal{I}_{ij}$, the transition is to ascribe to multipartite correlations.

In the present paper, we investigate the critical behavior of entanglement measures underlying the above classification for an extended Hubbard model at $T=0$ exactly solvable in one dimension. This is achieved by using appropriate measures of two-point and/or multipartite quantum correlations developed recently in quantum information, with particular emphasis on negativity, \cite{vidalvernerneg} concurrence, \cite{Woo} and entanglement ratio. \cite{VerER} In particular, in order to generalize the latter to qudit systems, the correlation ratio is introduced. Also, the relation between critical exponents at the transitions and the scaling behavior of the entanglement measure at those critical points is studied.

The paper is organized as follows. In Sec. \ref{Sec:model}, we introduce the model and its exact solution; we also derive the one- and two-site reduced density matrices. In Sec. \ref{Sec:measures}, we describe different measures of bipartite and/or multipartite correlations for qubit and qudit systems. In Sec. \ref{Sec:results}, we present and discuss the results obtained for the various measures at the different metal-insulator-superconducting transitions which characterize the model. Finally, in Sec. \ref{Sec:concl}, we summarize our main conclusions.

\section{The bond-charge extended Hubbard model}
\label{Sec:model}

The model is described by the following Hamiltonian:
\bae\label{ham_bc}
    H_{BC} =  &-& \sum_{<i,j>\sigma}[1 - x (n_{i \bar{\sigma}}+n_{j \bar{\sigma}})]
    c_{i \sigma}^\dagger c_{j \sigma} -\mu\sum_{i\sigma}n_{i\sigma}\nonumber\\
    & & + u \sum_i \left( n_{i \uparrow}-\frac{1}{2}\right) \left(n_{i \downarrow}-\frac{1}{2}\right)   \, ,
\eae
where $c_{{i} \sigma}^\dagger$ and $c_{{i} \sigma}^{} \,$ are fermionic creation and annihilation operators on a one-dimensional chain of length $L$; $\sigma = \uparrow, \downarrow$ is the spin label, $\bar{\sigma}$ denotes its opposite, ${n}^{}_{j \sigma} = c_{j \sigma}^\dagger c_{j \sigma}^{}$ is the spin-$\sigma$ electron charge, and $\langle {i} , \, {j} \rangle$ stands for neighboring sites on the chain; $u$ and $x$ ($0\leq x \leq 1$) are the (dimensionless) on-site Coulomb repulsion and bond-charge interaction parameters; $\mu$ is the chemical potential, and the corresponding term allows for arbitrary filling.\\
\begin{widetext}
\begin{figure}[!h]
    \begin{center}
        \fbox{\includegraphics[height=7cm, width=7cm, viewport= 10 20 280 230,clip]{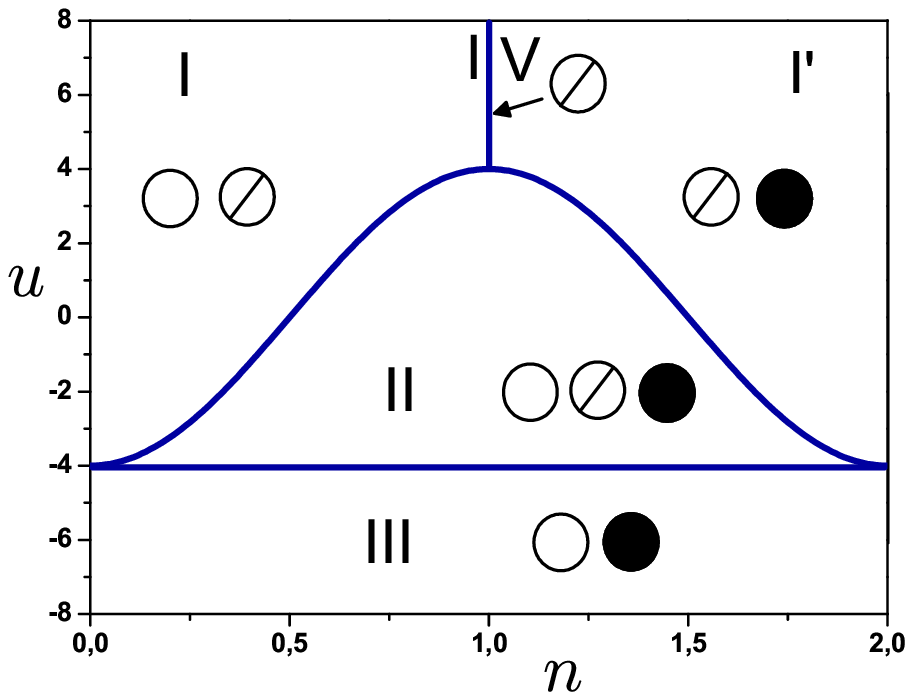}
        \includegraphics[height=7cm, width=7cm,viewport= 10 12 280 217,clip]{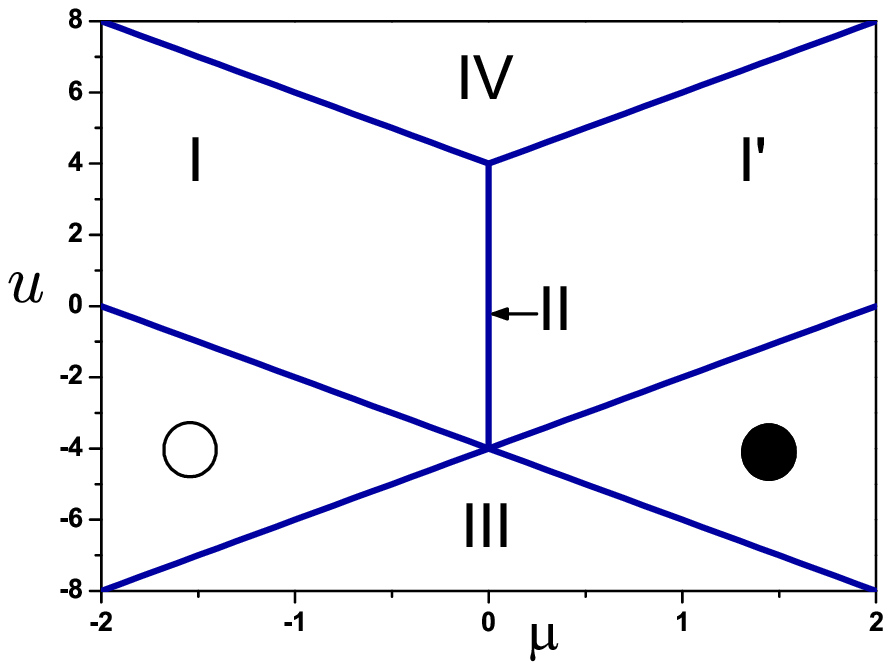}}
        \caption{Ground-state phase diagram of $H$. LEFT: $n$-$u$
         plane; empty circles stand for empty sites, slashed and full
         circles stand for singly and doubly occupied sites, respectively. RIGHT:
         $\mu$-$u$ plane.}\label{FGS1}
    \end{center}
\end{figure}
\end{widetext}
The model is considered here at $x=1$, in which case the number of doubly occupied sites becomes a conserved quantity and the role of spin orientation becomes irrelevant to many aspects: for an open chain, any sequence of spins in the chain cannot be altered by the Hamiltonian, whereas for periodic boundary conditions, only the sequences of spins related by a cyclic permutation can be obtained. In particular, the ground state turns out to be degenerate with the fully polarized state, and the system behaves as if at each site $i$ the local space had dimension $D_i=3$. In practice, both the Hamiltonian and the local vector space can be written in terms of the Hubbard-like projection operators $X^{\alpha\beta}_i\doteq\ket{\alpha}_i\bra{\beta}_i$, with local algebra $X^{\alpha\beta}_i X^{\gamma\delta}_i = \delta_{\beta\gamma}X^{\alpha\delta}$ and nonlocal (anti-)commutation relations given by
\be\label{HubProj}
    X^{\alpha\beta}_i X^{\gamma\delta}_j=(-)^{(\alpha+\beta) (\gamma+\delta)}
    X^{\gamma\delta}_j X^{\alpha\beta}_i \; , \; i\neq j \, ;
\ee
here $\alpha=0,1,2$, $\ket{0}_i\equiv\ket{\mbox{vac}}_i$ is the local vacuum, $\ket{1}_i\doteq X^{10}_i \ket{0}_i$ is the singly occupied state (with odd parity), and $\ket{2}_i\doteq X^{20}_i \ket{0}_i$ is the doubly occupied state.
More precisely, as far as the ground state is concerned, the model Hamiltonian in the one-dimensional case can be fruitfully written as
\bae
    H= &-& \sum_i \left (X^{10}_i X^{01}_{i+1}-X^{21}_i X^{12}_{i+1}
    + \mbox{H.c.}\right ) +u\sum_i X^{22}_i\nonumber \\
    &-&\left(\mu+{u\over 2}\right) \sum_i \left(X^{11}_i+2 X^{22}_i\right) \; . \label{hred}
\eae
In this form, $H$ provides the full spectrum of $H_{BC}$ at $x=1$ for open boundary conditions and its full ground-state phase diagram for both open and periodic boundary conditions.

\subsection{Spectrum and ground-state phase diagram}

The physics of the system described by $H$ is basically that of $N_s=\langle\sum_i X^{11}_i\rangle$ spinless fermions which move in a background of $L-N_s$ bosons, of which $N_d=\langle\sum_i X^{22}_i\rangle$ are doubly occupied sites and the remaining are empty sites. Both $N_s$ and $N_d$ are conserved quantities, and determine the total number of electrons $N=N_s+2 N_d$.\\
The situation may be understood in the formalism developed by Sutherland in Ref. \onlinecite{SUT}. We can say that, apart from constant terms, $H$ acts as a permutator of just two {\it Sutherland species} (SSs), the $N_s$ fermions, and the $L-N_s$ bosons. In practice, empty and doubly occupied states ---though different as physical species--- belong to the same SS, since the off-diagonal part of the Hamiltonian cannot distinguish between them. It is only the constant term counting doubly occupied sites which depends on the actual value of $N_d$.

The eigenstates are easily worked out, \cite{AA,DOMO} and read
\be
    |\psi (N_s,N_d)> = \mathcal{N} (\eta^\dagger)^{N_d} \tilde X^{10}_{k_1}
    \cdots \tilde X^{10}_{k_{N_s}}\ket{\mbox{vac}} \, ; \label{psi}
\ee
the result also holds at finite $L$ if periodic boundary conditions are chosen in Eq. (\ref{hred}). Here
\[
    \mathcal{N}=\Bigl[(L-N_s-N_d)!/(L-N_s)!N_d!\Bigr]^{1/2}
\]
is a normalization factor; $\tilde X^{10}_k$ is the Fourier transform of the Hubbard projection operator $X^{10}_j$, i.e., $\tilde X^{10}_k= \sum_j {1\over \sqrt{L}}\exp(i { \pi\over L} j k) X^{10}_{j}$. Moreover, $\eta^\dagger =\sum_{i=1}^L X^{20}_i$ is also known as the eta operator, commuting with $H$; $(\eta^\dagger)^{N_d}$ creates $N_d$ pairs (or doubly occupied sites).

The actual ground state $\ket{\psi_{GS}(N_s,N_d)}$ is chosen among the eigenstates in Eq. (\ref{psi}) by requiring that $N_s$ and $N_d$ minimize the corresponding eigenvalue
\[
    E(N_s,N_d)= -2  \frac {\sin \left(\pi\frac{N_s}{L}\right)}
    {\sin \left(\frac{\pi}{L}\right)}+u N_d-\mu (N_s+2 N_d)\, .
\]

In Fig. \ref{FGS1}, we report on the left the ground-state phase diagram in the $n$-$u$ plane (with $n=N/L$ average per-site filling); on the right part, the same diagram in the $\mu$-$u$ plane. The phase diagram presents various QPTs driven by parameters $u$ and $\mu$ (or $n$). Table \ref{table: TGS1} gives the range of parameters characterizing each phase in the thermodynamic limit ($L,N_\alpha \rightarrow \infty$, with $n_\alpha=N_\alpha/L$ finite). Each transition is characterized by a change in the number of on-site degrees of freedom (DOF) involved in the state. Phase IV has just one DOF per site since each site is singly occupied. It is an insulating phase, with charge gap $\Delta_c^{\mbox{IV}}=\mu_+-\mu_-= u-4$, where $\mu_+$ ($\mu_-$) is the energy cost for adding (removing) one electron. Phases I and I' (which is the particle-hole counterpart of phase I) have two on-site DOF: singly occupied sites and empty or doubly occupied sites respectively. This holds for phase III as well, where only empty and doubly occupied sites appear. Phase II is the only phase in which all three on-site DOF are involved.
\begin{widetext}
\begin{center}
\begin{table}[h]
    \begin{tabular}{lccc}
        \hline\hline
        Region of the phase diagram & $u$ & $\mu$ & GS energy \\
        \hline
        I: $n_s=n$ &
        $u>u_c(n)$ &
        $\mu=-2\cos{\pi n}-u/2$ &
        $-2/\pi\sin\left(\pi n\right)$ \\
      $\quad\quad\quad n_d =0$ & &  & \\

      I': $n_s=2-n$ &
        $u>u_c(n)$
&  $\mu=2\cos{\pi n}+u/2$ & $-2/\pi\sin\left(\pi n\right)+u (n-1)$ \\
      $\quad\quad\quad n_d =n-1$ & &  & \\

        &  & \\
        II: $n_s=1/\pi\arccos\left(-u/4\right)$ &
        $u\in [-4,u_c(n)]$
        &$\mu=0$ & $-2/\pi\sqrt{1-{\left(u/4\right)}^2}+u/2 [n-1/\pi \arccos
        (-u/4)]$\\
        $\quad\quad\;n_d=1/2 (n-n_s)$ & & & $ $\\
        III: $n_s=0$ & $u<-4$ & $\mu_\pm=\mp(2+u/2)$& $u n/2$ \\
       $\quad\quad\quad n_d=n/2$ & & & \\

        IV: $n_s=n=1$ & $u>4$ & $\mu_\pm=\mp(2-u/2)$ & $0$\\
       $\quad\quad\; n_d=0$ & & & \\
        \hline
        \hline
    \end{tabular}
\caption{Ground-state sectors and corresponding energies. Here $u_c(n)\doteq-4 \cos (\pi n)$. Note that the values limiting the range of $u$ and/or $\mu$ in each sector are the critical values for the transitions.} \label{table: TGS1}
\end{table}
\end{center}
\end{widetext}
Note that, as far as the relevant physics is concerned, this seems to be related to the number of on-site SSs characterizing the phase rather than the number of on-site DOF. In fact, phases I, I' and II ---which all have both the bosonic and the fermionic SSs--- fall in the Tomonaga-Luttinger class, since neither spin nor charge gap are present, whereas phase III, though characterized by empty and doubly occupied states, has just the bosonic SS; it is again insulating, with charge gap $\Delta_c^{\mbox{III}}=-u-4$.

Despite the above observation, phases I, I' differ form phase II since only  the latter is characterized by the occurrence of off-diagonal long-range order (ODLRO) and superconducting correlations (which also survive in phase III):
\be\label{ODLRO}
    \lim_{r\rightarrow\infty}\langle X^{20}_i  X^{02}_{i+r}\rangle = n_d(1-n_d)\, .
\ee
Note that ODLRO ---though not allowing real superconducting order at $x=1$ due to spin degeneracy, which implies the vanishing of spin gap \cite{AAG}--- is at the very root of superconducting order, which occurs at $x\neq 1$. \cite{ADMO}

Before discussing the various transitions in terms of the behavior of entanglement measures, let us recall some feature of each of them in terms of standard theory. First of all, since $N_d$ and $N_s$ are both conserved quantities, the transitions should be originated from level crossing. Indeed, they also occur at finite $L$. Nevertheless, none of them is of first order, since it can be easily checked from Table \ref{table: TGS1} that the first derivative of $E_{GS}$ is always smooth. Second, $N_s$ ($1-N_s$) and $N_d$ ($1-N_d$) can also be considered as order parameters for the transitions, since all of the QPTs occur in correspondence with the vanishing of one or both the above quantities. Moreover, all the three transitions I$\rightarrow$ IV, II$\rightarrow$ III and II$\rightarrow$ IV correspond to the opening of an insulating phase, characterized by a charge gap linear in $u$ (and in $\mu$). This implies that the product of the dynamical exponent $z$ and the critical exponent $\nu$ of the correlation length is 1 for all the three transitions. Furthermore, since $\rho=(d+z)\nu-1$, with $\rho$ exponent characterizing the first derivative of the free energy, we obtain that all of the three above transitions have $\nu=1/2$ and $z=2$. The situation is less clear at the transition II$\rightarrow$ I,I', since on the one hand no spin nor charge gap opens in both phases. On the other hand, it must be said that $N_d$, and consequently  $\eta$ pairs and ODLRO, vanish in correspondence with the transition. This is true as well for  the pairing gap $\Delta_P$, $\Delta_P=E(N+2)+E(N)-2 E(N+1)$. Indeed it can be seen that $\Delta_P=0$ in phase I, and $\Delta_P= u-u_c(n)<0$ in phase II, where $u_c(n)$ defines the critical line.

\subsection{Reduced density matrices}

The evaluation of the measures of correlation described in the following sections requires the manipulation of the single-site and dimer reduced density matrices, $\rho_i$ and $\rho_{ij}$. These can be obtained from the system density matrix in the ground state; the latter being defined as usual by $\rho_{GS}\doteq\ket{\psi_{GS} }\bra{\psi_{GS}}$. The reduced density matrix $\rho_i$ ($\rho_{ij}$) is then the trace of $\rho_{GS}$ with respect to all the DOF except those of the site $i$ (sites $i,j$). These matrices can be constructed in a simple way from the one- and two-point correlation functions, using the \emph{operator expansion} for the density matrix in terms of the Hubbard projectors (\ref{HubProj}); one, however, has to pay attention to the existing graded structure of the fermionic algebra, \cite{ANMO} (see Appendix \ref{Ap: trick}). In particular, the one-site reduced density matrix can be written as
\[
    \rho_{i}=\mbox{Tr}_{L/i} \left(\rho_{GS}\right)=\sum_{\alpha,\beta=0,1,2}
    q_{\alpha\beta} X^{\alpha\beta}_i,
\]
where $ q_{\alpha\beta} =\langle\psi_{GS}|X_i^{\alpha\beta}|\psi_{GS}\rangle$, while the two-site reduced density matrix reads
\[
    \rho_{ij} = \mbox{Tr}_{L/\{ij\}}\left(\rho_{GS}\right) =
    \sum_{\alpha,\beta,\gamma,\delta=0,1,2}q_{\alpha\beta\gamma\delta}
    X^{\alpha\beta}_i X^{\gamma\delta}_j,
\]
with $q_{\alpha\beta\gamma\delta} = \langle\psi_{GS}|X_i^{\alpha\beta}X_j^{\gamma\delta} |\psi_{GS}\rangle$. Below we report the results for $\rho_i$ and $\rho_{ij}$ (the detailed derivation of the calculations for the dimer case can be found in Appendix \ref{Ap: calcoli}).

When expressed in terms of the basis $\{\ket{0}_i,\ket{1}_i,\ket{2}_i\}$, $\rho_i$ is diagonal in all the regions of the phase diagram:
\bae
    \rho_i=\mbox{diag } \{1-n_s-n_d,n_s,n_d\}\quad ,\label{rhoi}
\eae
whereas with respect to the basis $\{\ket{00},\ket{01},\ket{\mbox10},\ket{11}, \ket{12},\ket{21},\ket{02},\ket{20}, \ket{22}\}$, $\rho_{ij}$ reads
\be
\rho_{ij}= \left( \begin{array}{ccccccccc}
D_1  & 0      & 0    & 0    & 0      & 0    & 0  & 0  & 0    \\
0    & O_1    & O_2  & 0    & 0      & 0    & 0  & 0  & 0    \\
0    & O_2^*  & O_1  & 0    & 0      & 0    & 0  & 0  & 0    \\
0    & 0      & 0    & D_2  & 0      & 0    & 0  & 0  & 0    \\
0    & 0      & 0    & 0    & P_1    & P_2  & 0  & 0  & 0    \\
0    & 0      & 0    & 0    & P_2^*  & P_1  & 0  & 0  & 0    \\
0    & 0      & 0    & 0    & 0      & 0    & Q  & Q  & 0    \\
0    & 0      & 0    & 0    & 0      & 0    & Q  & Q  & 0    \\
0    & 0      & 0    & 0    & 0      & 0    & 0  & 0  & D_3  \\
\end{array} \right)\, .\label{rhoij}
\ee
Here
\[
    \begin{array}{cclccclcc}
        D_1 & = & P_{ij} (1-c)\frac{1-c-\ep}{
        1-\ep}\, ,\quad & & \;  O_2 &=& \Gamma_{ij}(1-c)\, ,\\
        D_2 & = & n_s^2- |\Gamma_{ij}|^2  \, ,\quad & & \; P_1 &=&
        c\left( 1-n_s -P_{ij}\right )\, ,\\
        D_3 &=& c \frac{c-\ep}{1-\ep} P_{ij}\, ,\quad
        & & \; P_2 &=& c \Gamma_{ij}\, ,\\
        O_1 &=& \left(1-n_s-P_{ij}\right) (1-c)\, ,\quad & & \; Q &=&
        \frac{c(1-c)}{1-\ep}P_{ij}\, ,\\
    \end{array}  \label{entries}
\]
with $c=n_d/(1-n_s)$, $P_{ij}=(1-n_s)^2-|\Gamma_{ij}|^2$, $\ep= c/L$, and
\[
    |\Gamma_{ij}|={1\over L}\frac{\sin (n_s\pi |i-j|)}{\sin({\pi\over L} |i-j|)}\, .
\]

\section{Measures of Entanglement}
\label{Sec:measures}

The theory of quantum information has provided the study of complex quantum phenomena, such as QPTs, of new and well-defined tools. In many cases, these tools have been used to describe the (critical) behavior of relevant many-body systems. In general, the fact that a critical behavior can be spotted by appropriate measures of entanglement should not be a surprise from the point of view of the Landau theory, since any measure of entanglement can be expressed as a unique functional of first derivatives of the ground-state energy. \cite{WSL} Nevertheless, the use of more advanced tools could provide new interesting features difficult to extract from standard theory. For instance, in Ref. \onlinecite{AGM} we have described how, by using the appropriate measures of bipartite correlations, it is possible not only to fully describe the phase diagram of some model, but also to discriminate the role of two-point from multipartite entanglement at each of the QPTs the system undergoes.

\subsection{Separating two-point from multipartite entanglement at QPTs}\label{SSQ2QS}

In this section we briefly  recall the method used in Ref. \onlinecite{AGM} where we were interested in the existing correlations between (a) the single site $i$ and the rest of the system, and (b) the generic site $i$ and a generic site $j\neq i$.

Since the full system is in a pure (ground) state, the amount of quantum correlations between a single site and the rest of the system is measured by the Von Neumann entropy of $\rho_i$:
\be
    \mathcal{S}_i=\mathcal{S}(\rho_i)=-\sum_{j=1}^D \lambda_j \log_2 \lambda_j \quad , \label{VNE}
\ee
where $\lambda_j$, $j=1,\dots,D$, are the eigenvalues of the reduced density matrix $\rho_i$.

The total correlations (quantum and classical) between two sites $i,j$ are captured by the quantum mutual information:
 \be
    \mathcal{I}_{ij}=\mathcal{S}(\rho_{i})+\mathcal{S}(\rho_{j})- \mathcal{S}(\rho_{ij})
     \, , \label{QMI}
\ee
where $\mathcal{S}(\rho_{ij})$ is the two-site Von Neumann entropy, $\mathcal{S}(\rho_{ij})= \sum_{j=1}^{D^2}\tilde \lambda_j \log_2 \tilde\lambda_j $, and  $\tilde \lambda_j, j=1,\dots,D^2$, are the eigenvalues of $\rho_{ij}$.

To resume, we have the following situation. On the one hand, the single site is quantum correlated with the rest of the system in two possible way: via two-point correlations ($Q2$), when it is ``individually" correlated with some/all the other sites, and via multipartite correlations ($QS$), when it is connected through $n$-point quantum correlations. On the other hand, the mutual information  allows one to evaluate all the correlations connecting two sites; the latter can be of a quantum nature, the already mentioned $Q2$, and/or of a classical nature ($C2$).

While the single-site entanglement $\mathcal{S}_i$ is not able to distinguish between $Q2$ and $QS$, the quantum mutual information $\mathcal{I}_{ij}$ is not able to distinguish between $Q2$ and $C2$. Nevertheless, in Ref. \onlinecite{AGM} we have shown that a comparison of the singular behavior of $\mathcal{S}_i$ with that of $\mathcal{I}_{ij}$ allows one to discriminate whether a QPT is ascribed to $Q2$ or $QS$ correlations. In fact, whenever the singular behavior exhibited by $\mathcal{S}_i$ is due to $Q2$ correlations, the \textit{same} type of singular behavior is necessarily displayed  by $\mathcal{I}_{ij}$ as well, since it also contains $Q2$ correlations.

\subsection{Measuring two-point entanglement}

The task of measuring quantum correlations between two given sites $i$ and $j$ has a simple solution when $i$ and $j$ are two-level systems (qubit) in terms of the {\it concurrence}. \cite{Woo} Even when $i$ and $j$ are arbitrary qudit, the quantification of entanglement can be carried on by means of the \emph{negativity}; \cite{vidalvernerneg} the latter being a lower bound for concurrence. \cite{Verstraete2}

\subsubsection{Concurrence}

The concurrence was first introduced in Ref. \onlinecite{Woo} and, for the case of two qubits, it is directly related to the entanglement of formation. In order to evaluate the concurrence one has to first manipulate the two-qubit density matrix $\rho_{ij}$ and find $\tilde{\rho_{ij}}=\rho_{ij} \sigma_y \bigotimes\sigma_y \rho_{ij}^*  \sigma_y \bigotimes\sigma_y $, where $\rho_{ij}^*$ is the elementwise complex conjugate of $\rho_{ij}$. The concurrence can then be written as
\be
    \mathcal{C}_{ij}=\mathcal{C}(\rho_{ij})=\mbox{max}\{0, \lambda_1-\lambda_2-\lambda_3-\lambda_4 \}\; ,
    \label{Conc}
\ee
where the $\lambda_i$'s are the square roots of the eigenvalues of $\tilde{\rho_{ij}}$ taken in decreasing order.

\subsubsection{Negativity}

Another measure of bipartite quantum correlations $Q2$ is the negativity,
\be
    \mathcal{N}_{ij}=\mathcal{N}(\rho_{ij})=(\|\rho_{ij}^{T_i}\|_1-1)/(d-1)\; ,
    \label{neg}
\ee
where $\rho_{ij}^{T_i}$ is the partial transposition with respect to the subsystem (site) $i$ applied on $\rho_{ij}$, and $\|O\|_1\doteq \mbox{Tr}\sqrt{ O^\dagger O}$ is the trace norm of the operator $O$. $\rho_{ij}^{T_i}$ can have negative eigenvalues $\mu_n$ and the negativity can also be expressed as $\mathcal{N}(\rho_{ij})=|\sum_n \mu_n|$. Although the negativity is not a perfect measure of entanglement, \cite{convexroofneg} since it fails to signal the entanglement in the subset of mixed states called partial positive transpose states, it gives important bounds for quantum information protocols, i.e., teleportation capacity and asymptotic distillability.

One would reasonably expect the measures of two-point entanglement to exhibit the same singular behavior of $\mathcal{S}_i$ and $\mathcal{I}_{ij}$ when the transitions are ascribed to $Q2$ correlations. This is not always the case.\cite{AGM} To understand such an unexpected feature, we shall explore in more detail the behavior of both  $\mathcal{N}_{ij}$ and $\mathcal{C}_{ij}$ when $r=|i-j|$ is varied in proximity of the QPTs dominated by $Q2$ correlations.

\subsection{Multipartite entanglement measurements}\label{MultEnt}

The case in which the singular behavior of $\mathcal{S}_i$ is ascribed to $QS$ correlations can also  be treated with the described bipartite measures in a simple, though not complete way. The only thing that one can say is that when the $QS$ correlations enter into play, the same singular behavior should not be displayed by $\mathcal{I}_{ij}$ or by $\mathcal{N}_{ij}$, since both measures regard only two-point correlations.

We now proceed to review the measures of multipartite entanglement useful for our analysis.

\subsubsection{Residual entanglement: The tangle}

The idea of {\it residual entanglement} was first introduced in Ref. \onlinecite{CoKuWotang} where the case of a three-qubit system in a pure state $\ket{\psi_{ABC}}$ was studied. The basic ideas are as follows:\\

(i) The concurrence for a {\it two-qubit pure state}  reduces to
\bae
    \mathcal{C}_{A, B}=2\sqrt{\mbox{det }\rho_{A}}\; .
\eae

(ii) Once  a {\em focus} qubit is chosen, in this case $A$, the following inequality holds for a {\it three-qubit pure state}:
\bae
    4 \mbox{det }\rho_A \ge \mathcal{C}^2_{AB}+\mathcal{C}^2_{AC}\; .
\eae

(iii) In the case of a {\it three qubit pure state}, the subsystem constituted by the pair $(B,C)$ is four dimensional, but only two of these dimensions can be used to express the state; in other words, both the reduced density matrices $\rho_A$ and $\rho_{BC}$ have only two nonzero eigenvalues. This fact leads one to interpret $2\sqrt{\mbox{det}\rho_A}$ as the concurrence between $A$ and $(B,C)$ and thus to rewrite the above inequality as
\be
     \mathcal{C}^2_{A,(BC)} \ge \mathcal{C}^2_{AB}+\mathcal{C}^2_{AC}\; .
     \label{ineqC3}
\ee
The last result says that the entanglement that the focus qubit $A$ can establish with each of the other qubits separately is bounded by the entanglement that it can globally establish with them. (The latter is a property  that is not satisfied by the entanglement of formation.) The definition of residual entanglement for a {\it three-qubit pure state}, or {\em tangle},  can be introduced in  the following unique way on the basis of the above results and of the fact that they do not depend on the focus qubit chosen:
\bae
    \tau_{ABC}=\mathcal{C}^2_{A,(BC)} - \mathcal{C}^2_{AB}-\mathcal{C}^2_{AC}\;.
\eae
Due to the permutation invariance, this quantity properly measures at least an aspect of three-qubit entanglement: the {\em three-way entanglement}.

\subsubsection{Entanglement ratio}

An example of the use of the tangles for the exploration of QPTs in spin systems is given in Ref. \onlinecite{VerER}. There, in order to detect the relevance of the two-point entanglement versus the $n$-way entanglement ($n > 2$), the ``CKW conjecture'' was assumed, i.e., the conjecture that the inequality (\ref{ineqC3}) can be extended to states of an arbitrary number of qubits: \cite{CoKuWotang}
\bae\label{ineqCsq}
    \tau_1=\mathcal{C}^2_{A,(BC..N)} \ge \mathcal{C}^2_{AB}+\mathcal{C}^2_{AC}+\dots
    +\mathcal{C}^2_{AN}=\tau_2 \, .
\eae
Note that for spin systems all the concurrences can be easily evaluated due to the qubit nature of the subsystems. In Ref. \onlinecite{VerER}, starting from the above conjecture, the authors define the {\em entanglement ratio} $E_R$ as follows:
\be
    E_R=\tau_2/\tau_1 <1 \; . \label{EntRat}
\ee
The more the ratio decreases, the more $QS$ correlations are relevant with respect to $Q2$ ones. Recently the CKW conjecture has been rigorously proven in Ref. \onlinecite{Verstraete3}.

Generalizations of the above results were carried out in Refs. \onlinecite{WongChristangle} and \onlinecite{YuSongntangle}, where the authors provided a {\em bipartite} entanglement measure for the case of {\em arbitrary dimensions} of the subsystems by defining, even in this case, the notion of tangle. The latter construction is, in general, difficult to apply, since the determination of the generalized bipartite concurrence requires application of optimization processes. In our case, this implies that $E_R$ can be easily applied only in the phase where the local Hilbert space is of a two-qubit kind, i.e., in phase I and III.

In order to overcome this problem and study the transitions II $\rightarrow$ I,III,IV, we will make use of a different kind of ratio that allows one to compare the {\em total} two-point correlations with the total correlations (quantum) of a single site with respect to the others. This can be done by substituting in the definition of the entanglement ratio: (i) the sum of the squares of two-site concurrences $\tau_2$ with the sum of quantum mutual information $\mathcal{I}_{ij}$, i.e.,
\be
    \tilde{\tau}_2=\sum_{j=1}^{L-1}  \mathcal{I}_{ij}\; ;
\ee
(ii) the linear entropy $\tau_1$ with the single-site entanglement $\tilde{\tau}_1= \mathcal{S}_i$. The new ratio, termed correlation ratio, reads
\be
    C_R =\tilde{\tau}_2/\tilde{\tau}_1.\label{CorrRat}\; .
\ee
The fact that quantum correlations cannot be freely shared by many object is encoded in the so-called monogamy principle demonstrated in Ref. \onlinecite{Verstraete3}, which is at the base of the definition of $\tau_2$. Classical correlations $C2$ are not required to satisfy this principle; hence the sum of the mutual information of a given site with the remaining of the lattice is, in general, not bounded. Such a feature, however, does not affect the change of $C_R$ at QPTs. $C_R$ compares the two-point correlations ($Q2$ + $C2$) of the site $i$ with its total correlations, that, in our case, are purely quantum ($Q2$ + $QS$). As we shall show, it is a useful tool to characterize the phase transitions II $\rightarrow$ I,III,IV in terms of two-point versus shared correlations.

\section{Results and discussion}
\label{Sec:results}

In this section we first derive in the $\mu-u$ setting the results achieved by the method presented in Ref. \onlinecite{AGM} about the two-point and/or multipartite nature of the entanglement involved at each transition for the model Hamiltonian $H_{BC}$. We then deepen the analysis by employing the measures described above [Eqs. (\ref{neg}), (\ref{EntRat}) and (\ref{CorrRat})] at the same transitions.

The method described in Sec. \ref{SSQ2QS} classifies the type of entanglement involved at a given QPT by direct comparison of the derivatives of $\mathcal{S}_i$ and $\mathcal{I}_{ij}$. In order to evaluate these two quantities, the eigenvalues of the reduced density matrices $\rho_i$, and $\rho_{ij}$  are needed. While the on-site reduced density matrix (\ref{rhoi}) is already diagonal, the two-site density matrix (\ref{rhoij}) is block diagonal and, in its diagonal form, reads
\be
    \tilde \rho_{ij} =\mbox{diag}\{D_1,\mathcal{O}_+,
    \mathcal{O}_-,D_2,\mathcal{P}_+, \mathcal{P}_-,\mathcal{Q}_+
    ,\mathcal{Q}_-,D_3\},
\ee
where, in the thermodynamic limit (TDL),
\begin{eqnarray}
    \mathcal{O}_\pm &=& (1-c)\Big[n_s(1-n_s)+|\Gamma_{ij} |\left(|\Gamma_{ij}|\pm
     1\right)\Big],\\
    \mathcal{Q}_+   &=& 2 c (1-c) \Big[(1-n_s)^2-|\Gamma_{ij}|^2\Big]\;,\\
    \mathcal{Q}_-   &=& 0,\\
    \mathcal{P}_\pm &=& c \Big[n_s(1-n_s) +|\Gamma_{ij} |\left(|\Gamma_{ij}|\pm
    1\right)\Big].
\end{eqnarray}
In Table \ref{tableresults} we summarize the behavior of the various functionals evaluated at the transition points. In the last column, the transition is labeled as $Q2$ whenever the divergencies displayed by $\partial_x S_i$ and $\partial_x \mathcal{I}_{ij}$ are of the same type. In the other cases, i.e., when only $\partial_x S_i$ displays a divergency, the transitions are labeled $QS$. As already discussed in Sec. \ref{SSQ2QS}, the two groups reflect the relevance and the role of the $Q2$ and $QS$ correlations at the different transitions. The detailed analysis carried out in the following sections not only fully confirms the existence of these two groups, but also gives evidence of a further unexpected critical phenomenon occurring at each of the transition of $Q2$ type.

A first consideration about Table \ref{tableresults} is in order. As already mentioned, we consider the derivatives with respect to $u$ and $\mu$,  which are the quantities that parametrize the Hamiltonian (\ref{ham_bc}). At variance with the study in the $u-n$ setting developed in Ref. \onlinecite{AGM}, the transition I,I' $\rightarrow$ IV is here described by $\partial_\mu S_i$ and $\partial_\mu \mathcal{I}_{ij}$. This allows us to properly include it in the $Q2$ group.
\begin{widetext}
    \begin{center}
    \begin{table}[h]
    \begin{tabular}{lccccccc}
        \hline
        \hline
            &   $\partial_x\mathcal{S}_{i}$  &   $\partial_x\mathcal{I}_{ij}$   &
            $\partial_x\mathcal{N}_{ij}$
            &   $\partial_x \mathcal{C}_{ij}$   & $R_\mathcal{N}$ & $\partial_x E_R$
            & Ent\\
        \hline
            I,I' $\rightarrow$ IV $(x=\mu)$
            & $1/\sqrt{|\mu-\mu_c|}$     &   $1/\sqrt{|\mu-\mu_c|}$   &   $ -1/\pi^2$ &
            $1/\sqrt{|\mu-\mu_c|}$       & $1/\sqrt{|\mu-\mu_c|}$
            & $1/\sqrt{|\mu-\mu_c|}$   & $Q2$ \\

            II $\rightarrow$ I,I' $(x=u)$
            & $\log(u_c-u)$&  f &  f &  & f  &   &  $QS$   \\

         II $\rightarrow$ III $(x=u)$
            &  $1/\sqrt{u-u_c}$  & $1/\sqrt{u-u_c}$  &  $1/(2 \pi^2)$ &  &
            $1/\sqrt{u-u_c}$     & & $Q2$\\

         II $\rightarrow$ IV $(x=u)$
            &  $1/\sqrt{u_c-u}$ & $1/\sqrt{u_c-u}$  & $-1/(4 \pi^2)$ &   &
            $1/\sqrt{u_c-u}$ &   & $Q2$\\
        \hline
        \hline
    \end{tabular}
        \caption{\small{Behavior of $R_{\mathcal{N}}$ and the evaluated partial
         derivatives at the various QPTs (left column): the critical values $u_c$ and $\mu_c$ can be inferred from Table \ref{table: TGS1}. ``f" stands for finite value.}}
        \label{tableresults}
    \end{table}
    \end{center}
\end{widetext}

\subsection{Two-point entanglement at $Q2$ transition points}

In this section, we analyze the behavior of the two-point correlations at each of the $Q2$ transitions. We first proceed with the computation of the negativity (\ref{neg}). At variance with the concurrence, the negativity can be used even when the local subsystem's Hilbert space has dimension greater than 2, i.e., region II. We apply the partial transposition to $\rho_{ij}$, and then we proceed with its diagonalization. $\rho_{ij}^{T_i}$ is block diagonal and the only nondiagonal subblock reads
\bae\label{rhotrasp}
    \left(
    \begin{array}{ccc}
        D_1  & O_2  & Q    \\
        O_2  & D_2  & P_2  \\
        Q    & P_2  & D_3  \\
    \end{array}\right)\;.
\eae
The negative eigenvalues of $\rho_{ij}^{T_i}$ coincide with the negative eigenvalues of this subblock. In the TDL, the only possibly negative eigenvalue reads
\be
    \lambda_{-} = \frac{1}{2}\left[aP_{ij}+D_2- \sqrt{\left[aP_{ij}-D_2\right]^2
    +4a|\Gamma_{ij}|^2}\right]\;,
\ee
with $a=c^2+(1-c)^2$.
A straightforward calculation shows that
\bae
    \mathcal{N}_{ij}=\left\{
    \begin{array}{ll}
        -\lambda_-\;,\qquad &
        \gamma_-^2<\Gamma_{ij}^2<\gamma_+^2\\
        0 &  \mbox{otherwise }\; ,
    \end{array}\right.  \label{negij}
\eae
where $\gamma_{\pm}^2 =  \Big[1-n_s(1-n_s)\Big]\pm \sqrt{1-2n_s(1-n_s)}$.

Quite interestingly, the result shows that the region characterized by nonvanishing negativity, i.e., $\Gamma_{ij}^2\in[\gamma_-^2,\gamma_+^2]$, depends only on $n_s$. Such a result reveals that the presence of $Q2$ correlations  in the TDL is deeply connected with the presence of the fermionic Sutherland specie: when the latter is absent, all negativities go to zero, whereas when $n_s\neq 0$, the relative number of empty or doubly occupied sites does not influence the presence of two-point entanglement.

\subsubsection{Concurrence versus Negativity}\label{SecConcNeg}

\begin{figure}[h]
  \begin{centering}
    \fbox{\includegraphics[height=6cm, width=8cm, viewport= 10 10 260 215,clip]{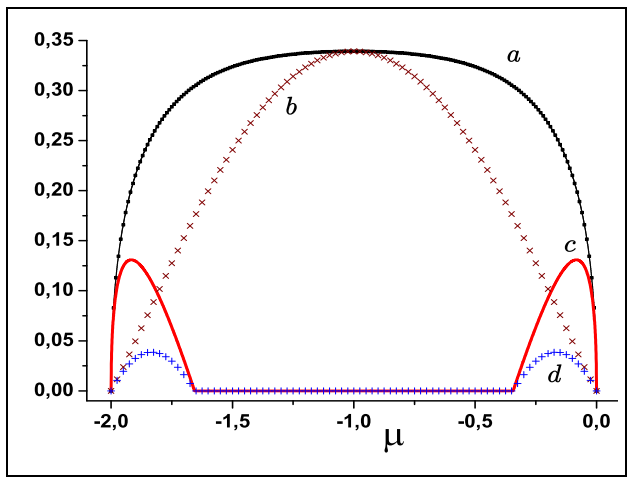}}
    \caption{Region I,  $u=4$. The curves $a$ and $c$ are the concurrencies $\mathcal{C}_{1}$ and $\mathcal{C}_2$, respectively, while the curves $b$ and $d$ are the negativities $\mathcal{N}_1$ and $\mathcal{N}_2$, respectively.}
    \label{Fig:Neg-Reg-I}
 \end{centering}
\end{figure}

Before proceeding in our analysis of the behavior of $Q2$ correlations at the transition points, it is useful to compare results obtained through negativity (\ref{negij}) with those obtained with concurrence (\ref{Conc}) in the regions where the latter can be evaluated, i.e., regions I and III. As we shall see, even though it has been proven that both are measures of entanglement for qubit systems, the comparison shows that, in general, they have different behaviors and derivatives. As far as phase I is concerned, we have that the concurrence is given by
\be
    \mathcal{C}^{I}_{ij}=2\mbox{ max}\left\{0,|\Gamma_{ij}|-\sqrt{\left[(1-n)^2-
    |\Gamma_{ij}|^2\right] (n^2-|\Gamma_{ij}|^2)}\right\} \;, \label{ConcI}
\ee
whereas specializing Eq. (\ref{negij}) to the case of region I, we obtain
\bae
    \mathcal{N}_{ij}^{I} &=& \mbox{max}\left\{0,\frac{1}{2}\left[(1-n)^2+n^2-2|
    \Gamma_{ij}|^2 \right.\right.\nonumber\\
    &&\left.\left.-\sqrt{(1-2n)^2+4|\Gamma_{ij}|^2 }\right]\right\}\; .
\eae
Figures \ref{Fig:Neg-Reg-I} and \ref{FNeg} show that, as expected, both measures are nonvanishing in the same intervals [see Eq. (\ref{negij})]. Indeed, for $r=|i-j|>1$, they start to differ from zero in correspondence with the same values of $\mu$. When both concurrence and negativity are nonzero, their behavior differs in at least two relevant aspects. First, apart from the case $r=1$, the two quantities reach their maximum in correspondence with two different values of $\mu$. Such a feature is compatible with the fact that, in general, the two measures provide a different ordering of the states. \cite{Verstraete2} Second, they differ in the behavior of their derivatives with respect to $\mu$. In particular, while at transition I$\rightarrow$ IV the derivative of concurrence does display the correct diverging behavior ($\partial_\mu \mathcal{C}_{ij} \approx 1/\sqrt{|\mu_c-\mu|}$),  $\partial_\mu \Nij$ does not display any divergence.
As for region III, in the TDL both $\mathcal{N}_{ij}$ and $\mathcal{C}_{ij}$ are always zero. The behavior of the two quantities significantly differs if finite-size effects are included. Indeed, to first order in $1/L$ and for all $|i-j|$, the concurrence reads
\bae\label{ConcIII}
    \mathcal{C}_{ij}^{\mbox{III}}=\left\{\begin{array}{ll}
        {1\over L}\;, \qquad & n_d\neq 0\\
        0 \qquad &  \mbox{otherwise ,}
    \end{array} \right.
\eae
while the negativity is \cite{AGM}
\be\label{negregIII}
    \mathcal{N}_{ij}^{\mbox{III}} = \frac{n_d (1-n_d)}{n_d^2+(1-n_d)^2} {1\over L} \; .
\ee

\subsubsection{Divergence of the entanglement range}\label{DivEntRan}

The fact that $\mathcal{N}_{ij}$ differs from zero at different values of $\mu$  depending on $r=|i-j|$ allows one to identify the range of negativity
\[
    R_\mathcal{N}(u,\mu) =\{r \ \  | \ \ \mathcal{N}_{i,i+r}\neq 0\,\,
    \wedge\, \mathcal{N}_{i,i+r+1}=0  \}\;,
\]
i.e., the maximum distance $r$ at which the Negativity is non-vanishing at fixed $u$, $\mu$. We find that $R_\mathcal{N}$ is always finite except in the two following situations: (i) when $n_s\rightarrow 0$ (transition II $\rightarrow$ III) and (ii) when $n_s\rightarrow 1$ (transitions I $\rightarrow$ IV and II $\rightarrow$ IV), in which $R_\mathcal{N}$ diverges. In particular, $R_\mathcal{N}$ remains finite in correspondence with the two transitions II $\rightarrow$ I, at which $n_s\rightarrow n\neq 1$ [and correspondingly $u\rightarrow u_c(n)$]. In these cases, only the nearest-neighbor negativity is always positive, i.e., $R_\mathcal{N}\ge 1$. The condition that fixes the value of $R_\mathcal{N}$ is again $\Gamma_{ij}^2 = \gamma_-^2$.

\begin{figure}[h]
    \begin{centering}
        \fbox{\includegraphics[height=6cm, width=4.1cm,viewport= 10 10 270 240,clip]{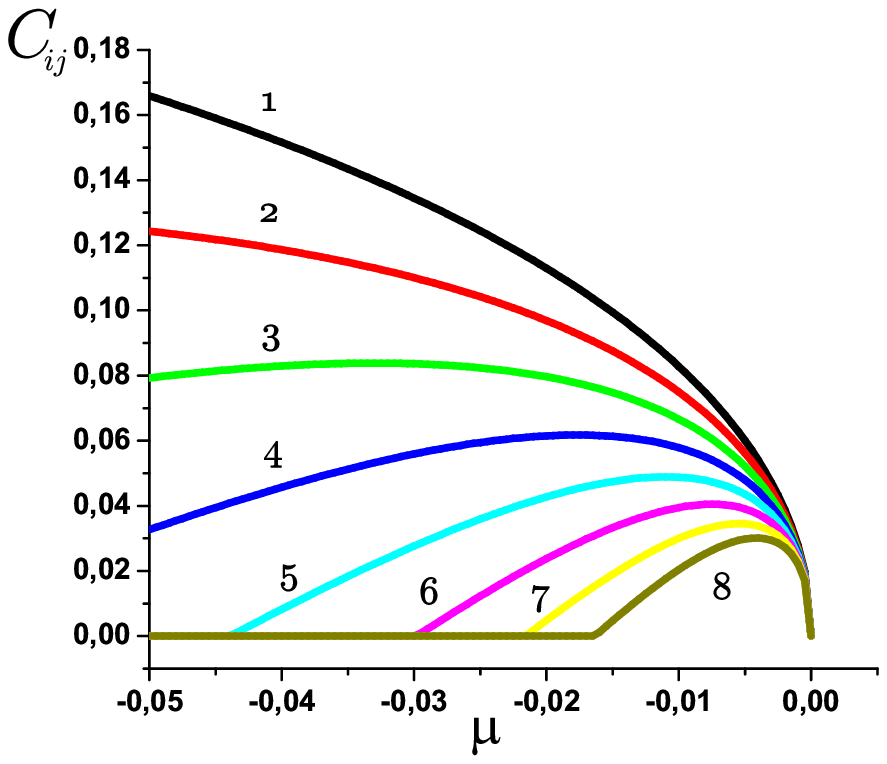}
        \includegraphics[height=6cm, width=4.1cm,viewport= 10 10 340 240,clip]{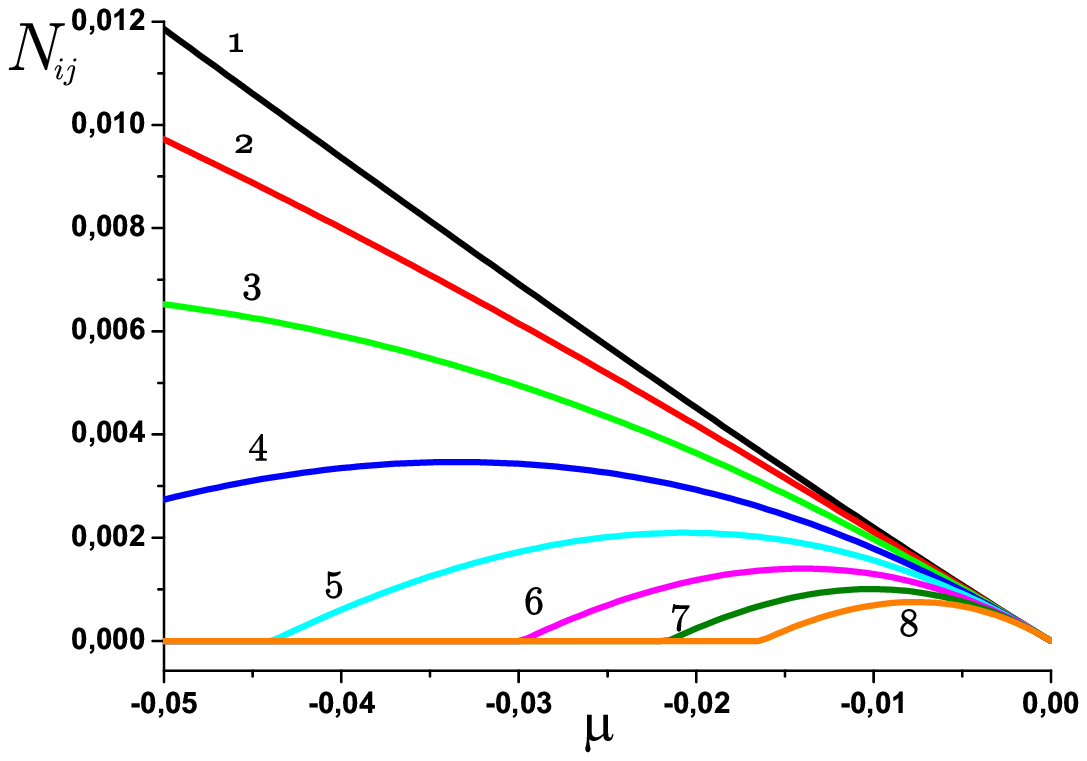}}
        \caption{Region I,  $u=4$.  $\mathcal{N}_{ij}$ (right) and $\mathcal{C}_{ij}$
        (left) for $r=1,..,8$.}
    \label{FNeg}
    \end{centering}
\end{figure}

The entanglement range allows one to better characterize the difference between $Q2$ and $QS$ transitions. In the latter, the generic site $i$ is correlated ---via two-point quantum correlations--- only with a finite number of neighboring sites, whereas, at $Q2$ transitions, the two-point quantum correlations begin to spread along the chain and, at the critical point, two arbitrarily distant sites are quantum correlated. This latter case is shown in Fig. \ref{FNeg} for the transition I $\rightarrow$ IV, at which $\mathcal{C}_{ij}$ can be evaluated as well.
We can further characterize the spreading of the correlations by analytically exploring  the scaling behavior of $R_\mathcal{N}$ at the transitions. We have that
\bae\label{negrange}
    R_{\mathcal{N}} \approx \left\{\begin{array}{ll} \frac{R_0}{n_s}\;, \qquad &
        n_s\rightarrow 0 \\
        \frac{R_0}{1-n_s}\;, \qquad &  n_s\rightarrow 1
    \end{array} \right.\;,
\eae
where $R_0\approx 0.44$ is the solution of
\be
    \frac{\sin{\pi R_0}}{\pi R_0}=\frac{1}{\sqrt{2}}\; .
\ee
The exponents $\nu_c$ characterizing the divergence of $R_{\mathcal{N}}$ at $u_c$ and $\mu_c$  for the various transitions are easily worked out from Eq. (\ref{negrange}) by recalling that $n_s$ is a function of $u$ and $\mu$ (see Table \ref{table: TGS1}). Quite interestingly, at all the three $Q2$ transitions, we have
\be
    \nu_c=\frac{1}{2}=\nu \; .
\ee

A similar type of behavior was already studied for a spin model  \cite{AmVerETr} where the notion of \emph{entanglement transition} was introduced. In that case, the divergence of the entanglement range is observed for the concurrence at some specific point of the phase diagram for which the ground state becomes factorized, and apparently no QPT takes place. We recognize such a feature for the model discussed here only in part. In fact, here all the entanglement transitions occur in correspondence with QPTs; moreover, while phase IV is indeed characterized by a factorized structure (the ground state being singly occupied at each site), phase III is not, since the ground state in this phase is a superposition of empty and doubly occupied states distributed over the whole chain. This observation suggests the conjecture that a factorized structure with respect to Hilbert space appearing in the ground state is a {\it sufficient} but not necessary condition for the occurrence of an entanglement transition.
The conjecture could also be generalized in terms of bipartite entanglement:
an entanglement transition occurs if and only if the new phase is a \emph{two-point entanglement free} one. In this sense the factorized state of phase IV and the genuine multipartite  ground state of phase III are equivalent. Moreover, at least in our model, it is equivalent the way in which the system destroys all correlations (IV) or build genuine multipartite ones (III).

\subsection{Two-point versus multipartite entanglement at QPTs}

In order to explore the role of multipartite entanglement at the various transitions, an ideal tool would be the entanglement ratio $E_R$ (\ref{EntRat}), which, as explained, provides a direct measure of the relative role of $Q2$ and $QS$: a decreasing (increasing) entanglement ratio in proximity of a QPT means that $QS$ ($Q2$) correlations are more relevant to the transition. According to  Eq. (\ref{ineqCsq}), $\tau_2$ is properly measured through concurrence. This implies that only the transitions in which the system is of qubit nature (namely, the I$\rightarrow$ IV transition) can be explored through $E_R$, whereas in region II we used the correlation ratio.

\subsubsection{Entanglement ratio}

\begin{figure}[h]
    \begin{centering}
        \fbox{\includegraphics[height=6cm, width=8cm,viewport= 10 15 320 225, clip]{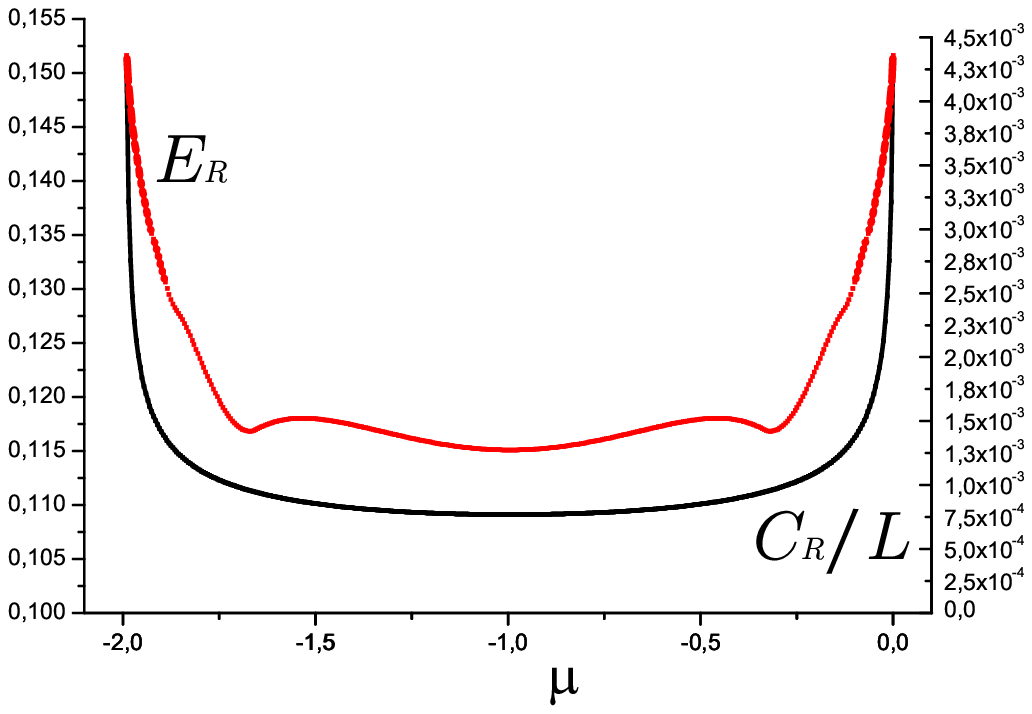}}
        \caption{Region I,  $u=4$. Entanglement ratio $E_R$ and normalized correlation ratio $C_R/L$ with $L=1000$.}
    \label{Fig:Ratio-Reg-I}
    \end{centering}
\end{figure}

We start exploring the behavior of $E_R$ in phase I, where $\mathcal{C}_{ij}$ is defined, and, in particular, at the transition I $\rightarrow IV$. As far as $\tau_1$ is concerned, from Eq. (\ref{ineqCsq}) we have
\be\label{tau1I}
    \tau^{I}_1=4 \mbox{ det } [\rho_i^{I}]= 4 n(1-n)\; .
\ee
As for $\tau_2$, the two-site concurrence was given in Eq. (\ref{ConcI}); still, the evaluation of the sum of the $(L-1)$ $\Cij^2$ in Eq. (\ref{ineqCsq}) requires some attention. In fact, $\mathcal{C}_{ij}$ depends on the distance $r$ between the two sites; one has to first evaluate the sum
\be
    \tau_2(N_s,L) =\sum_{r=1}^{L-1} \mathcal{C}^2_{r}(N_s,L)
\ee
and then the TDL $\lim_{N_s,L \rightarrow \infty}\tau_2(N_s,L)$. The numerical evaluation of $E_R$ for a sufficiently large $r$ is reported in Fig. \ref{Fig:Ratio-Reg-I}.
The latter clearly shows that, as expected, in the vicinity of transition I $\rightarrow$ IV ($\mu\rightarrow 0$) the $Q2$ correlations rapidly increase with respect to $QS$ ones. We notice that interestingly the derivative of $E_R$ diverges again with $|\mu|^{-1/2}$ at the transition.

We pass to explore $\tau_2$ in region III, which is again of qubit nature. Since such a region is characterized by the presence of just $\eta$ pairs, which have an intrinsic multipartite nature, we expect  $E_R$ to vanish there. As far as $\tau_1$ is concerned, we have that
\be\label{tau1III}
    \tau^{\mbox{III}}_1=4 \mbox{ det } [\rho_i^{\mbox{III}}]= 4 n_d(1-n_d)\, ,
\ee
whereas for $\tau_2$ one can see that, since $\mathcal{C}_{ij}^{(\mbox{III})}\sim 1/L$ [see Eq. (\ref{ConcIII})] is independent of $r=|i-j|$ and vanishing in the TDL, $\tau_2=\sum_j \mathcal{C}_{ij}^2=0$. Hence, the entanglement ratio correctly indicates that \emph{the only relevant entanglement is multipartite} (i.e., $n$-way entanglement with $n \ge 3$).

\subsubsection{$QS$ correlations in region II}

As a general fact, in region II both $Q2$ and $QS$ correlations are present. In particular, the $QS$ transitions ---which according to Table \ref{tableresults} are II $\rightarrow$ I,I' at fixed $u$--- should be characterized by some change in multipartite entanglement $QS$. Such a hypothesis has a first strong confirmation in the fact that at these transitions $\eta$ pairs [and ODLRO, see Eq. (\ref{ODLRO})] disappear. Indeed, it has been shown in Ref. \onlinecite{vedraleta} that $\eta$ pairs do carry multipartite entanglement, thus disappearing at these transitions. Furthermore, the behavior of $Q2$ correlations is here radically different from the one they display at $Q2$ transitions. Actually, as already seen in Sec. \ref{DivEntRan}, the entanglement range is finite at any filling $n<1$ for $u\leq u_c(n)$; moreover, both $R_\mathcal{N}$ and $\Nt_{ij}$, and their derivatives remain finite in the same regime. At $n=1$ and $u=4$ (i.e., at transition II$\rightarrow$ IV), besides $QS$ also $Q2$ correlations enter to play a role. In fact, while $\eta$ pairs disappear, $R_\mathcal{N}$ becomes infinite and an entanglement transition takes place. Since the analysis of the previous section has shown that $R_\Nt$ has the same divergence of $\mathcal{S}_i$, we infer that the role of $Q2$ correlations is dominant at this transition. In order to confirm such a scheme, we now use the correlation ratio previously introduced.

\subsubsection{Correlation ratio}

We aim at obtaining an indicator of the relative weight of $Q2$ correlations with respect to $QS$ ones in region II. Since $\mathcal{I}_{ij}$ keeps track of the change of $Q2$ correlations between $i$ and $j$ at transition points, we expect the correlation ratio $C_R$ (\ref{CorrRat}) to capture such a desired feature.

We first consider the behavior of $C_R$ in region I, where it can be compared with the standard entanglement ratio $E_R$; this is shown for sufficiently large $L$ by the dashed line in Fig. \ref{Fig:Ratio-Reg-I}. In correspondence with the transition II $\rightarrow$ IV, $C_R$ correctly reproduces the qualitative behavior of $E_R$, i.e., the relative weight of $Q2$ correlations rapidly increases.
\begin{figure}[h]
    \begin{centering}
        \fbox{\includegraphics[height=6cm, width=8cm, viewport= 5 15 280 230,clip]{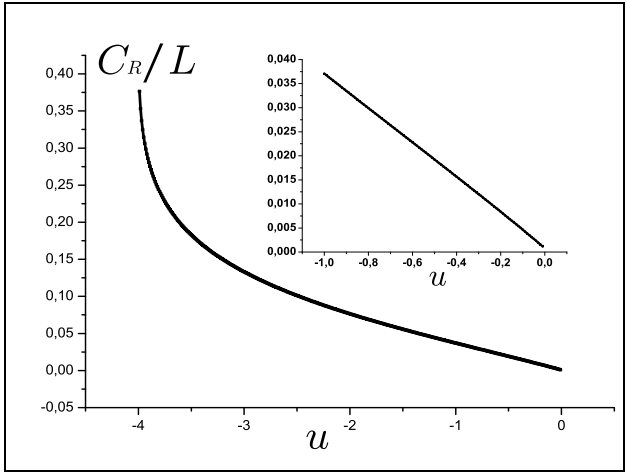}}
        \caption{Region II, $n=1/2$  $u\in[-4,0]$. Normalized correlation ratio $C_R/L$ with $L=1000$. Inset: zoom of the transition II $\rightarrow$ I.}
    \label{Fig:Ratio-Reg-IIa}
    \end{centering}
\end{figure}
\begin{figure}[h]
    \begin{centering}
        \fbox{\includegraphics[height=6cm, width=8cm, viewport= 5 180 280 400, clip]{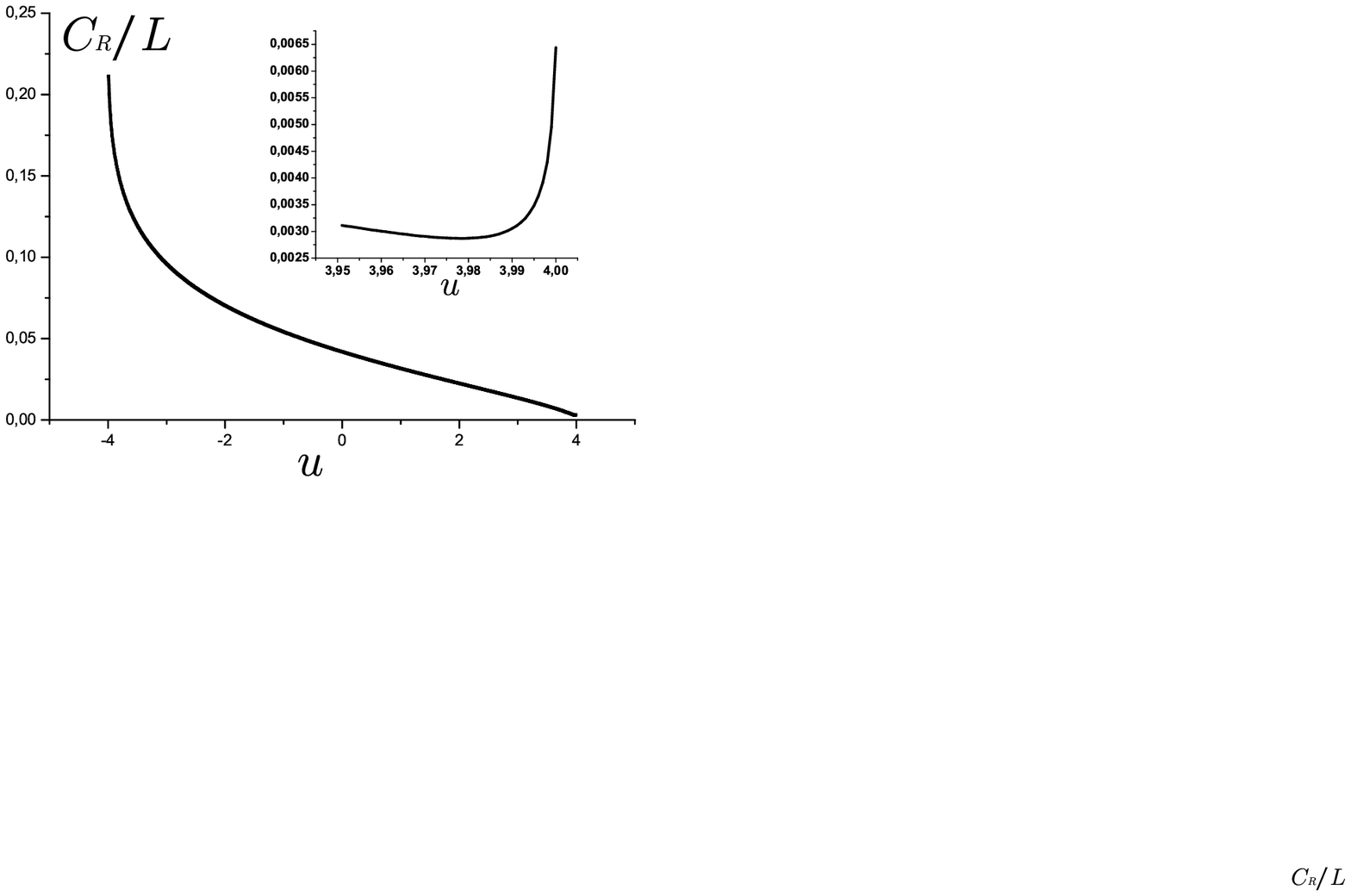}}
        \caption{Region II, $n=1$,  $u\in[-4,4]$.  Normalized correlation ratio $C_R/L$ with $L=1000$. Inset: zoom of the transition II $\rightarrow$ IV.}
    \label{Fig:Ratio-Reg-II}
    \end{centering}
\end{figure}
In Figs. \ref{Fig:Ratio-Reg-IIa} and \ref{Fig:Ratio-Reg-II} we report $C_R(u)$ in region II at two different values of $n$ ($n=1/2$ and $n=1$). At both values $C_R$ rapidly increases in proximity of $u=-4$ (transition II $\rightarrow$ III). The behavior of $C_R$ is quite different in the two cases in correspondence with the upper critical point. Indeed, for $n=1/2$ (transition II $\rightarrow$ I, Fig. \ref{Fig:Ratio-Reg-IIa}), it goes to zero with a clear linear dependence on $u_c-u$, reminiscent of the behavior of the pairing gap $\Delta_P$, whereas for $n=1$ (transition II $\rightarrow$ IV), after decreasing in almost the whole region, $C_R$ rapidly increases for $u\rightarrow 0$.

These features are in accordance with the considerations exposed in the previous section. Whenever $C_R$ increases at the transition, this implies that $Q2$ correlations are increasing with respect to $QS$, and hence the transition should be of $Q2$ type. On the other hand, it is only at the transition of $QS$ type (transition II $\rightarrow$ I) that $C_R$ vanishes, meaning that $QS$ correlations overcome $Q2$ ones.

\subsection{Entanglement away from QPTs}

Apart from transition points, we can spot the areas in the different regions where the $QS$ correlations prevail with respect to $Q2$ correlations from the direct study of $\mathcal{S}_i$, $\mathcal{N}_{ij}$. In Region III, $QS$ correlations prevail everywhere since $S_i$ is different from zero and all $\Nij$ are vanishing. Let us recall that one has genuine multipartite entanglement whenever both $S_i\neq 0$ and two-point entanglement is zero. Here it is due to the presence of $\eta$ pairs, which is also captured by two-point classical correlations. In fact, it turns out that $\mathcal{I}_{ij}^{(\mbox{III})}=\mathcal{I}_\infty^{(\mbox{III})} = 2n_d(1-n_d) \ \ \forall j$, with $\mathcal{I}_\infty\doteq \lim_{|i-j|\rightarrow \infty} \mathcal{I}_{|i-j|}$. All pairs of sites are equally correlated as two infinitely distant sites. Interestingly this property is directly related to the presence of ODLRO in that the total amount of correlations is simply proportional to it.

\begin{figure}[h]
    \begin{centering}
        \fbox{\includegraphics[height=6cm, width=8cm,viewport= 17 18 260 220, clip]{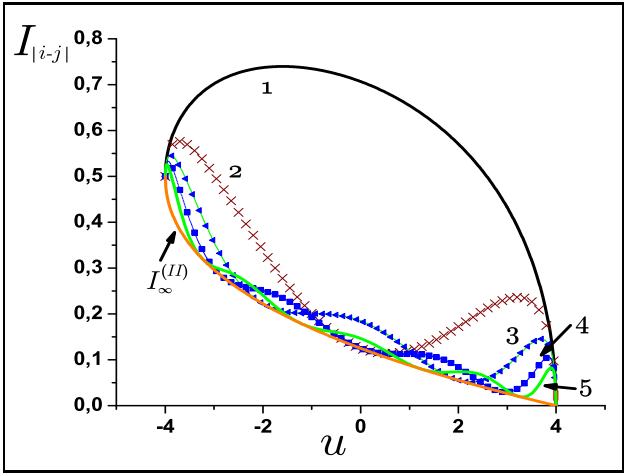}}
        \caption{Region II, $n=1$ $u\in[-4,4]$.
        Mutual information $\mathcal{I}_{ij}$, $r=1,..,5$.}
        \label{Fig: Mutual reg II}
    \end{centering}
\end{figure}

In region I, $QS$ correlations prevail away from transition points since the entanglement ratio has a minimum (see Fig. \ref{Fig:Ratio-Reg-I}). Contextually, only the nearest-neighbor negativity is nonzero, and $S_i$ is maximum; moreover, $\mathcal{I}_\infty^{(I)}=0$. The same qualitative behavior holds inside region II as well, except that $\mathcal{I}_\infty^{(II)}\neq 0$ in the whole region except at the transition II $\rightarrow$ I, as can be seen from the dashed line in Fig. \ref{Fig: Mutual reg II}. This is related with the fact that $\eta$ pairs are present in region II as well. Quite interestingly, the contribution of singly occupied and doubly occupied sites to two-point correlations seems to simply add in quantum mutual information. Indeed, one could check that $\mathcal{I}_{ij}^{(\mbox{II})}\approx \mathcal{I}_{ij}^{(I)}+\mathcal{I}_\infty^{(\mbox{II})}$.

To resume, we observe that an infinite range of two-point correlations in proximity of a transition is a signal of a $Q2$-driven QPT, in which case $R_{\mathcal{N}}$ also diverges. Far from transition, the same infinite range is implied
whenever $\mathcal{I}_\infty\neq 0$ and it is thus a signal of the existing ODLRO,
in our case related to $\eta$ pairs.
\begin{figure}[h]
    \begin{centering}
        \fbox{\includegraphics[height=6cm, width=8cm, viewport= 15 15 270 230,clip]{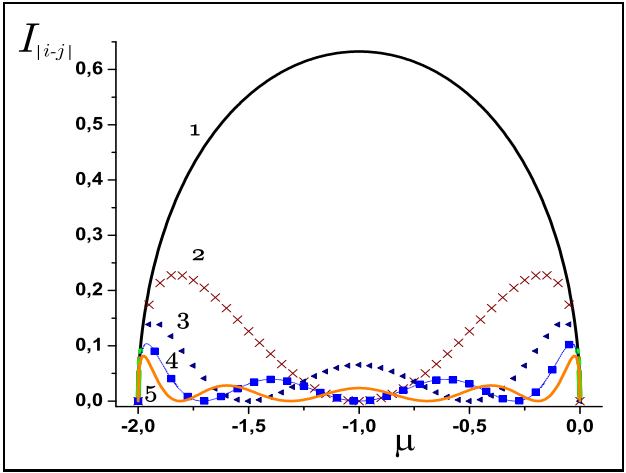}}
        \caption{Region I,  $u=4$. Mutual information $\mathcal{I}_{ij}$, $r=1,..,5$.}
        \label{Fig: Mutual reg I}
    \end{centering}
\end{figure}

\section{Conclusions}
\label{Sec:concl}

In this paper, we analyzed the rich phase diagram of the one-dimensional bond-charge extended Hubbard model at $T=0$ by means of various measures of bipartite and/or multipartite correlations. All the computed measures are capable of reproducing the known phase diagram in terms of singularities; moreover, at each transition the critical exponent of the correlation length is shown to coincide with the scaling exponent of the divergent quantities, when evaluated.

The knowledge of one- and two-site Von Neumann entropies allows one to distinguish the quantum phase transitions (QPTs), in which the role of two-point quantum correlations ($Q2$) is relevant to  those in which multipartite quantum correlations ($QS$) are determinant.

The systematic analysis of the appropriate measures of entanglement at the different transitions and phases made it possible to better characterize both $Q2$ and $QS$ transitions. Contextually, a different estimator which can be computed for qudit systems has been introduced: the \emph{correlation ratio}. The latter is an indicator of the relative weight of the two-point correlations with respect to the total (quantum) ones at the transitions. The analysis shows that $Q2$ transitions are characterized by a divergence in the range of negativity: when approaching the transition, two sites at arbitrary distance become quantum correlated. Such a feature is reflected in the behavior of the entanglement and/or correlation ratio. The latter increases at $Q2$ critical points with diverging derivative, clearly indicating an increasing relative weight of two-point quantum correlations with respect to multipartite ones. $QS$ transitions instead are characterized by a finite range of negativity and by a vanishing correlation ratio, indicating that multipartite quantum correlations dominate there. For our model, the correlated physical phenomenon is the disappearance of $\eta$ pairs.

Finally, we described the nature of the correlations within each region as well. For our model, the existence of two-point quantum correlations depends only on the presence of singly occupied sites. At the same time, the presence of doubly occupied sites witnesses the appearance of $\eta$ pairs and ODLRO, and multipartite entanglement carried by them.
At the level of two-point correlations ODLRO coincides with the finite value of quantum mutual information between infinitely distant sites.

In conclusion, the above analysis has widely clarified how to characterize the nature of quantum correlations involved at a QPT for an integrable correlated electron model. The scheme, in particular, allows one to gain from quantum mutual information insight on the behavior of both $Q2$ and $QS$ correlations at transition points for $T=0$. We expect the scheme to be straightforwardly applicable also in nonintegrable cases, both in one and in greater dimension. A first step in this direction has been achieved in one dimension by means of the numerical analysis in the nonintegrable case (\ref{ham_bc}). \cite{ADMO} It remains to be investigated how to modify the proposed scheme at $T\neq 0$, where also temperature-driven correlations play a major role; in particular, it is expected that they would compete with quantum ones in determining the behavior of quantum mutual information.

\begin{acknowledgements}
The authors gratefully acknowledge useful discussions with D. Larsson.
\end{acknowledgements}

\begin{appendix}

\section{Reduced Density Matrix expansion}\label{Ap: trick}

Let us recall the definition of the reduced density matrix $\rho_{ij}$:
\be\label{Ap:rho}
    \rho_{ij}=\mbox{Tr}_{L/ij}|\Psi_{GS}\rangle\langle\Psi_{GS}|\; ;
\ee
given that the pure state $|\Psi_{GS}\rangle$ can be written as
\be\label{Ap:basis}
    |\Psi_{GS}\rangle=\sum_{ab}C_{ab}|a\rangle|b\rangle\, ,\quad C_{ab}\in\mathbb{C}\,,
\ee
where $\{|a\rangle\}$ and $\{|b\rangle\}$ are the basis for the subsystems $\{ij\}$ and $\{L/ij\}$, respectively, the operator expansion for $\rho_{ij}$ is easily derived:
\begin{eqnarray}
    \rho_{ij} &=& \mbox{Tr}_{L/ij}\sum_{aa^{\prime}bb^{\prime}}
    C_{ab}C_{a^{\prime}b^{\prime}}^*|a\rangle|b\rangle \langle b^{\prime}|\langle
    a^{\prime}|\nonumber\\
    &=& \sum_{aa^{\prime}bb^{\prime}b^{\prime\prime}} C_{ab} C_{a^{\prime}b^{\prime}}^*
    \langle b^{\prime\prime} |a\rangle |b\rangle\langle b^{\prime}|\langle a^{\prime}|
    b^{\prime\prime}\rangle\nonumber\\
    &=& \sum_{aa^{\prime}bb^{\prime}b^{\prime\prime}} C_{ab} C_{a^\prime b^\prime}^*
    (-)^{\left(ab+a^{\prime}b^{\prime}\right)} |a\rangle\langle a^{\prime}|
    \delta_{bb^{\prime\prime}} \delta_{b^\prime b^{\prime\prime}} \nonumber\\
    &=& \sum_{aa^{\prime}b} C_{ab} C^*_{a^\prime b}(-)^{\left(a+a^{\prime}
    \right)b}|a\rangle\langle a^{\prime}|\nonumber\\
    &=& \sum_{aa^\prime} \mathcal{C}_{aa^{\prime}} |a\rangle\langle a^{\prime}|\;,
\end{eqnarray}
where
\be
    \mathcal{C}_{aa^\prime}=\sum_b C_{ab} C^*_{a^\prime b}=\langle\Psi_{GS}|a
    \rangle\langle a^\prime |\Psi_{GS}\rangle\;,
\ee
where $(-)^a$ takes into account the parity of the state $|a\rangle$. \cite{ANMO} It turns out that for our model $a+a^\prime=0$ for all $(a,a^\prime)$, due to the conservation of $N_s$ and $N_d$.

\section{Dimer Reduced Density Matrix Evaluation}\label{Ap: calcoli}

In this section, we schematically give the procedures to compute the dimer reduced density matrix. The mean value of the following operators give the diagonal elements of $\rho_{ij}$:
\bae\label{op-rho-d}
    \begin{array}{lll}
        X_i^{02}X_i^{20}X_j^{02}X_j^{20}, & X_i^{02}X_i^{20}X_j^{11}, & X_i^{11}X_j^{02}X_j^{20},\\
        X_i^{11}X_j^{11}, & X_i^{11}X_j^{22}, & X_i^{22}X_j^{11},\\
        X_i^{02}X_i^{20}X_j^{22}, & X_i^{22}X_j^{02}X_j^{20}, & X_i^{22}X_j^{22}.
    \end{array}
\eae
Due to the conservation of $N_s$ and $N_d$, the only off-diagonal elements that can be nonvanishing in some of the regions are the following ones (together with their Hermitian conjugates):
\bae\label{op-rho-fd}
    X_i^{10}X_j^{01},\; -X_i^{21}X_j^{12},\; X_i^{02}X_j^{20}.
\eae
The action of the latter is simply to permute the states on the two sites.

\subsubsection{Region I}\label{sec:I}

Since the ground state in region I is given by a superposition of states in which each site is empty or singly occupied and since $N_s=N_{tot}\doteq N$ and $X_i^{02}X_i^{20} \equiv [1-X_i^{11}]$ the only nonzero entries are given by
\begin{eqnarray}
    \langle X_i^{02}X_i^{20}X_j^{02}X_j^{20}\rangle_{\mbox{I}} &\equiv& \langle(1-X_i^{11})(1-X_j^{11})\rangle_{\mbox{I}},\nonumber\\
    \langle X_i^{02}X_i^{20} X_j^{11}\rangle_{\mbox{I}} &\equiv& \langle(1-X_i^{11})X_j^{11}\rangle_{\mbox{I}},\nonumber\\
    \langle X_i^{11}X_j^{02}X_j^{20}\rangle_{\mbox{I}} &\equiv&
    \langle X_i^{11}(1-X_j^{11})\rangle_{\mbox{I}},
\end{eqnarray}
and
\bae
    \langle X_i^{11}X_j^{11}\rangle_{\mbox{I}},\;
    \langle X_i^{10}X_j^{01}\rangle_{\mbox{I}}\;;
\eae
where $\langle \mathcal{O} \rangle_{\mbox{I}}\equiv \langle\Psi_{\mbox{I}}(N,L)| \mathcal{O}|\Psi_{\mbox{I}}(N,L)\rangle$.
We thus compute $\langle X_i^{11} \rangle_{\mbox{I}}$ using the expression of both $|\Psi_{\mbox{I}}(N,L)\rangle$ and  $X_i^{11}$ in terms of momentum operators
\bae\label{njdefk}
    \langle X_i^{11} \rangle_I = \frac{1}{L}\sum_{k,k'} \exp[i (k-k') j] \langle \tilde{X}_k^{10} \tilde{X}_{k'}^{01}\rangle_I= \frac{N}{L}\, ,
\eae
since the only nonvanishing terms of the sum in Eq. (\ref{njdefk}) are those for which $k-k'=0$.

The calculation of $\langle X_i^{10}X_j^{01}\rangle_{\mbox{I}}$ is analogous to the previous one; here we have the appearance of a phase factor:
\bae\label{cij}
    \langle X_i^{10}X_j^{01}\rangle_{\mbox{I}} = \frac{1}{L}\sum_m\exp[-i2\pi m(i-j)/L] \doteq \Gamma_{i-j}\; .
\eae
Such a phase factor has a different expression for $N$ even or odd:
\bae
    |\Gamma^E_{i-j}|=|\Gamma^O_{i-j}|=\frac{\sin\left( \frac{N}{L}\pi|i-j|\right)}{L\sin\left(\frac{1}{L}\pi|i-j|\right)}\;.
\eae
In the TDL, $\Gamma_{ij}=\Gamma^E_{ij}=\Gamma^O_{ij}$ and it can be computed by approximating the sum in Eq. (\ref{cij}) with the following integral:
\bae
    \Gamma_{ij} = \frac{1}{\pi}\int^{m \pi}_0\cos k(i-j) dk = \frac{\sin \left(n|i-j|\pi\right)} {\pi |i-j|}.
\eae
The previous calculations allow us to eventually compute
\bae
    \langle X_i^{11}X_j^{11}\rangle_{\mbox{I}} = \left(\frac{N}{L}\right)^2-\Gamma_{i-j} \Gamma_{j-i}.
\eae
The expressions of $D_1$, $D_2$, $O_1$, and $O_2$ in region I ($N_d=0$) follow from the collection of the previous results.

\subsubsection{Region III}\label{sec:III}

The ground state in region III is given by a superposition of states in which each site is empty or doubly occupied. The only nonzero entries are given by
\begin{eqnarray}\label{op-reg-III-1}
    \langle X_i^{02}X_i^{20}X_j^{02}X_j^{20}\rangle_{\mbox{III}},\;&& \langle X_i^{02}X_i^{20}X_j^{22}\rangle_{\mbox{III}},\nonumber\\
    \langle X_i^{22}X_j^{02}X_j^{20}\rangle_{\mbox{III}},&&
\end{eqnarray}
and
\bae\label{op-reg-III-bis}
     \langle X_i^{22}X_j^{22}\rangle_{\mbox{III}},\;\;
     \langle X_i^{02}X_j^{20}\rangle_{\mbox{III}}\;,
\eae
where $\langle \mathcal{O} \rangle_{\mbox{III}}\equiv \langle\Psi_{\mbox{III}}(N,L)| \mathcal{O}|\Psi_{\mbox{III}}(N,L)\rangle$.
The evaluation of Eq. (\ref{op-reg-III-1}) follows from the evaluation of Eq. (\ref{op-reg-III-bis}). A general strategy is used. In the case of $\langle X_i^{22}X_j^{22}\rangle_{\mbox{III}}$ one has to count the number of states in the superposition $|\Psi_{\mbox{III}}(N,L)\rangle$ whose sites $i$ and $j$ are doubly occupied:
\bae
    \langle X_i^{22}X_j^{22}\rangle_{\mbox{III}} = \mathit{N}_{\mbox{III}}(N_d) {L-2\choose N_d-2}
    =\frac{N_d(N_d-1)}{L(L-1)}\;.
\eae
In order to compute $\langle X_i^{02}X_j^{20}\rangle_{\mbox{III}}$ one has to count all the states whose site $i$ is doubly occupied and whose site $j$ is empty. This leads to the following result:
\bae
    \langle X_i^{02}X_j^{20}\rangle_{\mbox{III}} = \mathit{N}_{\mbox{III}}(N_d) {L-2\choose N_d-1}= \frac{N_d(L-N_d)}{L(L-1)}.
\eae
The expressions of $D_1$, $D_3$, and $Q$ in region III ($N_s=0$) follow from the previous results.

\subsubsection{Region II}\label{sec:II}

We start by noting that since $[X^{11}_i,\eta^\dagger]=0$, the operator $X^{11}_i$ does not affect the doubly occupied part of the ground state. Accordingly,
\begin{eqnarray*}
    \langle N_d| X^{11}_i|N_d\rangle_{\mbox{II}} &=& \frac{N_s}{L},\\
    \langle N_d| X^{11}_iX^{11}_j|N_d\rangle_{\mbox{II}} &=& {\left(\frac{N_s}{L}\right)}^2 -|\Gamma_{i-j}|^2\;,
\end{eqnarray*}
where the notation $\langle N_d|\mathcal{O}|N_d\rangle_{\mbox{II}}\equiv \langle\Psi_{\mbox{II}}(N_s,N_d,L) |\mathcal{O}|\Psi_{\mbox{II}}(N_s,N_d,L)\rangle $ will be useful in the following calculations.

We now compute $\langle N_d| X^{22}_j|N_d \rangle$. In a first step we make use of the following relations:

(i) $[X_j^{22},(\eta^\dagger)^{N_d}]= N_d(\eta^\dagger)^{N_d-1}X^{20}_j$,

(ii) $[X^{20}_j,(\eta^\dagger)^{N_d-1}]=0$,

(iii) $(\eta)^{N_d}=(\eta)^{N_d-1}\eta$.

The latter imply that
\begin{eqnarray*}\label{recur1}
    &&[\mathit{N}_{\mbox{II}}(N_d)]^2\langle \Psi_{\mbox{I}}|(\eta)^{N_d} X_j^{22}(\eta^\dagger)^{N_d} |\Psi_{\mbox{I}}\rangle\\
    &&=[\mathit{N}_{\mbox{II}}(N_d)]^2N_d\langle\Psi_{\mbox{I}}|(\eta)^{N_d} (\eta^\dagger)^{N_d-1}X^{20}_j|\Psi_{\mbox{I}}\rangle\\
    && =\frac{ [{\mathit N}_{\mbox{II}}(N_d)]^2 }{ [{\mathit N}_{\mbox{II}}(N_d-1)]^2}N_d^2\langle N_d-1|X^{02}_j X^{20}_j|N_d-1\rangle.\\
\end{eqnarray*}
If we now define ${\mathcal D}_{N_d-1} \doteq \langle N_d-1|X^{02}_j X^{20}_j|N_d-1\rangle$, we may write the following recursive equation:
\bae
    \mathcal{D}_{N_d}= 1- \frac{N_s}{L}-\frac{N_d}{L-N_s-N_d +1}\mathcal{D}_{N_d-1},
\eae
whose solution is ${\mathcal D}_{N_d}=\frac{L-N_s-N_d}{L}$. Thus, by collecting the above results, we have
\bae
    \langle N_d|X^{22}_j|N_d\rangle = \frac{ [{\mathit N}_{\mbox{II}}(N_d)]^2 }
    {[{\mathit N}_{\mbox{II}}(N_d-1)]^2}N_d^2\mathcal{D}_{N_d-1 }=\frac{N_d}{L}\;.
\eae
We now compute $\langle N_d|X^{11}_iX^{22}_j|N_d\rangle$ by resorting to the solution of a recursive equation. Since $[X^{22}_j, X^{11}_i]=0$, we can first apply the procedure used for $\langle N_d|X^{22}_j|N_d \rangle$ to obtain
\bae
    \langle N_d|X^{22}_j X^{11}_i |N_d\rangle=\frac{[{\mathit N}_{\mbox{II}}(N_d)]^2} {[{\mathit N}_{\mbox{II}}(N_d-1)]^2}(N_d)^2 \mathcal{E}_{N_d-1},
\eae
where $\mathcal{E}_{N_d-1}\doteq\langle N_d-1|X^{02}_j X^{20}_j  X^{11}_i|N_d-1\rangle$ satisfies the following recursive expression:
\begin{eqnarray}\label{sfuc}
    \mathcal{E}_{N_d-1} &=& \langle N_d-1|(1-X^{11}_j) X^{11}_i|N_d-1\rangle \nonumber\\
    && -\frac{[{\mathit N}_{\mbox{II}}(N_d-1)]^2}{[{\mathit N}_{\mbox{II}}(N_d-2)]^2} (N_d-1)^2\mathcal{E}_{N_d-2}\nonumber\\
    &=& \frac{N_s}{L}-\frac{N_s^2}{L^2}+|\Gamma_{i-j}|^2\nonumber\\
    && - \frac{N_d-1}{L-N_s-N_d+2}\mathcal{E}_{N_d-2}\;.
\end{eqnarray}
The solution of the latter is
\bae\label{E-N_d}
    \mathcal{E}_{N_d}=\left(1-\frac{N_d}{L-N_s}\right) \left[\frac{N_s}{L}\left(\frac{L-N_s}{L}\right)+ |\Gamma_{i-j}|^2\right]\;,
\eae
and, finally, we get
\bae
    \langle N_d|X^{22}_j X^{11}_i|N_d\rangle= \frac{N_d}{L-N_s}\left[\frac{N_s(L-N_s)}{L^2}+|\Gamma_{i-j}|^2\right]\;.
\eae
We now compute $\langle N_d|X_i^{22} X_J^{22}|N_d\rangle$. The same arguments used for Eq. (\ref{sfuc}) lead to
\bae
    \langle N_d|X^{22}_j X^{22}_i |N_d\rangle=\frac{[{\mathit N}_{\mbox{II}} (N_d)]^2}{[{\mathit N}_{\mbox{II}}(N_d-2)]^2}N_d^2(N_d-1)^2 \mathcal{F}_{N_d-2}\;,
\eae
where the function defined as $\mathcal{F}_{N_d}\doteq\langle N_d| X^{02}_j X^{20}_jX^{02}_i X^{20}_i|N_d\rangle$
satisfies the following recursive equation:
\bae
    \mathcal{F}_{m}=(\alpha-\alpha')\left(1-\frac{m}{\beta}\right) -\frac{m}{\beta-m+1}{\mathcal F}_{m-1}\;,
\eae
with
\bae
    \begin{array}{lcllcl}
        \alpha  & = & 1-\frac{N_s}{L}, & \beta   & = & L-N_s,\\
        \alpha' & = & \frac{N_s}{L}-\frac{N_s^2}{L^2}+| \Gamma_{i-j}|^2, & m & = & N_d.
    \end{array}
\eae
The latter recursive equation is solved by defining the auxiliary function $\mathcal{G}_m\doteq\frac{m}{\beta-m+1}\mathcal{F}_{m-1}\frac{1}{(\alpha-\alpha')}$ that obeys
\bae
    \mathcal{G}_m = \frac{m}{\beta}-\frac{m}{\beta-m+1}\mathcal{G}_{m-1}\, ,
\eae
whose solution is $G_m=-\frac{m(\beta-m)}{\beta(\beta-1)}$. Collecting the above results, we have that
\begin{eqnarray}
    &&\langle N_d|X^{22}_jX^{22}_i |N_d\rangle\nonumber\\ &&=\frac{N_d(N_d-1)\left[{\left(1-\frac{N_s}{L}\right)}^2- |\Gamma_{i-j}|^2\right]}{(L-N_s)(L-N_s-1)}\;.
\end{eqnarray}
The computation of $\langle N_d|X_i^{02}X_j^{20}|N_d\rangle$ is now straightforward, since it can be expressed in terms of the above defined $\mathcal{F}_m$:
\begin{eqnarray}
    &&\langle N_d|X_i^{02}X_j^{20}|N_d\rangle =\frac{ [{\mathit N}_{\mbox{II}}(N_d)]^2} {[{\mathit N}_{\mbox{II}}(N_d-1)]^2 }N_d^2\mathcal{F}_{N_d-1}\nonumber\\
    &=& \frac{N_d(L-N_s-N_d)\left[{\left(1-\frac{N_s}{L}\right)}^2 -|\Gamma_{i-j}|^2 \right]}{(L-N_s)(L-N_s-1)}.\nonumber\\
\end{eqnarray}
The same argument holds for $\langle N_d|X_i^{02}X^{20}_iX_j^{22}|N_d \rangle$:
\begin{eqnarray}
    \langle N_d|X_i^{02}X^{20}_i X^{22}_j|N_d \rangle &=& \frac{[\mathrm{N}_{\mbox{II}}(N_d)]^2}{[\mathrm{N}_{\mbox{II}}(N_d-1)]^2}N_d^2 \mathcal{F}_{N_d-1}\nonumber\\
    &=&\langle N_d|X_j^{02}X^{20}_i|N_d \rangle.
\end{eqnarray}
The previous steps allow us to easily evaluate $\langle N_d|X_i^{02}X^{20}_i X^{11}_j|N_d \rangle$ in terms of Eq. (\ref{E-N_d}):
\bae
    \langle N_d |X^{02}_i X^{20}_i X^{11}_j|N_d \rangle= \mathcal{E}_{N_d}.
\eae
We then compute $\langle N_d|X^{10}_iX^{01}_j|N_d\rangle$. Using the following relations:

(i) $[X^{10}_i X^{01}_j , (\eta^\dagger)^{N_d}]= -N_d X_i^{10} X^{21}_j(\eta^\dagger)^{N_d-1}$,

(ii) $(\eta)^{N_d}X^{21}_j=N_d(\eta)^{N_d-1}X^{01}_j$;
we obtain the recursive equation
\begin{eqnarray}\label{defdiG}
    \mathcal{L}(N_d) &\doteq& \langle N_d|X^{10}_i X^{01}_j|N_d\rangle\nonumber\\
    &=& \frac{N_d\mathcal{L}(N_d-1)}{N_d-L-N_s+1}+\Gamma_{i-j}\;,
\end{eqnarray}
whose solution is
\be
    \langle N_d|X^{10}_i X^{01}_j |N_d\rangle = \Gamma_{i-j}\frac{L-N_s-N_d}{L-N_s}\;.
\ee
The task of evaluating $\langle N_d|X_i^{12}X_j^{21}|N_d\rangle$ is simplified by observing that $[X_j^{10} X_i^{01}, \eta^\dagger] = [X_i^{21} X^{12}_j,\eta^\dagger]$; we obtain
\begin{eqnarray}
    \langle N_d|X^{21}_i X^{12}_j |N_d\rangle &=& (N_d)^2\frac{[\mathit{N}_{\mbox{II}}(N_d)]^2}{[\mathit{N}_{\mbox{II}}(N_d-1)]^2} \mathcal{L}(N_d-1)\nonumber\\
    &=& \frac{N_d}{L-N_s}\Gamma_{j-i},
\end{eqnarray}
where $\mathcal{L}(N_d-1)$ is given by Eq. (\ref{defdiG}).
\end{appendix}

\end{document}